# Investigating Some Diatomic Molecules Bounded by Two-Dimensional Isotropic Oscillator Plus Inverse Quadratic Potential in an External Magnetic field

O. J. Oluwadare[1*], E. O. Ilesanmi[1], T. O. Abiola[1], O. Olubosede[1], E. A. Odo[1], S. O. Ajibade[2] and K. J. Oyewumi[2]

[1] Department of Physics, Federal University Oye-Ekiti, PMB. 373, Ekiti State, Nigeria

[2,] Department of Physics, University of Ilorin, PMB 1515 Ilorin, Kwara State, Nigeria

# Investigating Some Diatomic Molecules Bounded by Two-Dimensional Isotropic Oscillator Plus Inverse Quadratic Potential in an External Magnetic field

O. J. Oluwadare[1*], E. O. Ilesanmi[1], T. O. Abiola[1], O. Olubosede[1], E. A.Odo[1], S. O. Ajibade[2] and K. J. Oyewumi[2]

[1] Department of Physics, Federal University Oye-Ekiti, PMB. 373, Ekiti State, Nigeria
[2,] Department of Physics, University of Ilorin, PMB 1515 Ilorin, Kwara State, Nigeria

*Email Addresses: oluwatimilehin.oluwadare@fuoye.edu.ng; emmanuel.ilesanmi.171020@fuoye.edu.ng; abiolato92@gmail.com, olusayo.olubosede@fuoye.edu.ng; ayodele.odo@fuoye.edu.ng; amtijsaheed@gmail.com; kjoyewumi66@unilorin.edu.ng*
*\*Corresponding author email: oluwatimilehin.oluwadare@fuoye.edu.ng*

**Abstract**

We investigate the nonrelativistic magnetic effect on the energy spectra, expectation values of some quantum mechanical observables and diamagnetic susceptibility for some diatomic molecules bounded by the Isotropic oscillator plus inverse quadratic potential. The energy eigenvalues and normalized wavefunctions are obtained via parametric Nikiforov-Uvarov method. The expectation values square of position $\langle r^2 \rangle$, square of momentum $\langle p^2 \rangle$, kinetic energy $\langle T \rangle$ and potential energy $\langle V \rangle$ are obtained by applying Hellmann-Feynman theorem and an expression for the diamagnetic susceptibility $X$ is also derived. Using the spectroscopic data, the low rotational and low vibrational energy spectra, expectation values and diamagnetic susceptibility $X$ for a set of diatomic molecules ($I_2$, $H_2$, CO, $HCl$) for arbitrary values Larmor frequencies are calculated. The computed energy spectra, expectation values and diamagnetic susceptibility $X$ were found to be more influenced by the external magnetic field strength and inverse quadratic potential strength $g$ than the vibrational frequencies and the masses of the selected molecules.

**Keywords:** Schrödinger equation; Isotropic oscillator plus inverse quadratic potential; magnetic field; parametric Nikiforov-Uvarov method; Hellmann-Feynman theorem; diatomic molecules;

**PACs numbers:** 03.65.-w; 03.65.Ca; 03.65.Ge; 03.65.Pm

## 1.0. Introduction

Several studies in quantum mechanics, solid state physics, condensed matter physics, nuclear physics, chemical physics, molecular physics and other related areas have proofed to an outstanding degree that potential models are very important models for stimulating atomic and molecular interaction since it is capable of predicting and describing the some behavior of atoms and molecules. It also provides an insight into the understanding of molecular spectra, vibrations and dynamics [1], [2], spin orbit interaction, relativistic corrections and diamagnetic susceptibility [3], [4], optical properties [5], [6], interband light absorption and interband optical transitions [7], [8], energy and relativistic effects in weakly bound nuclei [9], [10], [11], external magnetic field and/or Aharonov-Bohm flux fields [12], [13], [14], [15], [16], [17], interactions between magnetic and electric fields [18], thermal and/or thermodynamic properties [19], [20], [21], [22], [23], spin and pseudospin symmetries [24], two body effects [25], [26], [27], [28] among others.

One of the important potential models in this regards is the so-called isotropic oscillator plus inverse quadratic potential (IOPIQP) or anharmonic oscillator potential, which has been explored by some authors in both relativistic and nonrelativistic domain of quantum mechanics [29], [30], [31], [32], [33], [34]. An isotropic oscillator (three-dimensional harmonic oscillator) plus Inverse quadratic potential which may be defined [29], [30], [31], [32], [33], [34] as

$$V(r) = \frac{1}{2}\mu\omega^2 r^2 + \frac{g}{r^2}, \qquad (1)$$

where $g$ is the potential strength, $\mu$ represents mass of the vibrating molecules and $\omega$ is the angular frequency with which the molecules vibrate in the presence of magnetic field. Oyewumi and Bangudu [29] employed the hyperradial equation for isotropic harmonic oscillator plus inverse quadratic potential and presented the normalized hyperradial and hyperangular solutions and that the solutions depend on the dimension as well as the potential parameters. The hidden symmetries and thermodynamic properties for a Harmonic Oscillator plus an inverse square potential has been exposed by Dong *et al.*[30] while Arda and Sever reported the exact solutions of Schrodinger this potential within the framework of Laplace transform technique [31].

In the same vein, Abdelmadjid [32] also studied the exact nonrelativistic quantum spectrum systems for the isotropic harmonic oscillator plus inverse quadratic potential within the formalisms of both Boopp's shift method and standard perturbation theory in both noncommutativity two dimensional real space and phase (NC-2D: RSP) and presented the exact corrections for the spectrum and the associated noncommutative anisotropic Hamiltonian. Again, due to the unflinching interest, Abdelmadjid [33] looked into the effect of the both non commutativity of three dimensional space and phase on the Schrödinger equation with an isotopic harmonic oscillator plus inverse quadratic potential and reported the exact degenerated spectrum associated for noncommutative space and phase.

Furthermore, Dianawati *et al.* [34] investigated the Schrödinger equation with quantum deformation for three-dimensional harmonic oscillator plus inverse quadratic potential via hypergeometric method. The energy spectra which were calculated and visualized by Matlab R2013a were found to depend on the quantum deformation and quantum number.

It is in the light of relevance of this potential model that we are motivated to examined the two dimensional radial Schrödinger equation with isotropic harmonic oscillator plus inverse quadratic potential in an external magnetic field via parametric Nikiforov-Uvarov method, obtain the eigensolutions, discuss the behaviour of energy spectra, expectation values of some quantum mechanical observables and diamagnetic susceptibility for some selected diatomic molecules bounded by this interaction potential model.

The sensitivity of the bounded molecules in an external magnetic field (using arbitrary values of Larmor frequencies), low rotational and vibrational levels and inverse quadratic potential strength would be adequately investigated. The situation where Larmor frequency $\omega_L = 0$ implies the absence of external magnetic field whereas for Larmor frequencies $\omega_L > 0$ indicate the presence of external magnetic field. The case of low vibrational energy level ($n = 0, 1, 2, 3$), low rotational energy level ($m = 0, +1$) and inverse quadratic potential strengths ($g = 0, 1$) would be examined.

Other methods that can be used to solve the aforementioned bound state problems include: wave function ansatz method [13], asymptotic iteration method [16], formula method [17], Euler-Maclaurin approximation [23], Laplace transform technique [31], [35], supersymmetric approach [36], among others. In section 2, we give the review of the parametric Nikiforov-Uvarov method. Section 3 contains nonrelativistic eigensolutions, expectation values of some quantum mechanical observables and diamagnetic susceptibility of the isotropic harmonic oscillator plus inverse quadratic potential in external magnetic field. The results are discussed extensively in section 4 while concluding remarks are given in Section 5.

## 2.0. Review of parametric Nikiforov-Uvarov method

The parametric Nikiforov-Uvarov method is a straight forward, consistent and efficient analytical technique for analyzing second order linear differential equations arising from bound state problems. The choice of this method is due to the fact that, it has been proven to an outstanding degree and as well reported to give an excellent results in comparison with other methods in literature [12]. According to Nikiforov and Uvarov [37], the second order linear differential equation to reduces to the generalized equation of hypergeometric type [37], [38]. With an appropriate coordinate transformation $z = z(r)$, the equation take the form

$$\psi_{nl}''(z) + \frac{\bar{\tau}(z)}{\sigma(z)}\psi_{nl}'(z) + \frac{\bar{\sigma}(z)}{\sigma^2(z)}\psi_{nl}(z) = 0, \qquad (2)$$

where $\sigma(z)$ and $\bar{\sigma}(z)$ are polynomials, at most second degree and $\bar{\tau}(z)$ is a first degree polynomials.

To solve equation (2), one needs to break the wave function $\Psi(z)$ into parts as:

$$\psi_{nl}(z) = \phi(z)y(z), \qquad (3)$$

therefore equation (2) reduces to the hypergeometric type equation:

$$\sigma(z)y''(z) + \tau(z)y'(z) + \lambda y(z) = 0, \qquad (4)$$

where

$$\tau(z) = \bar{\tau}(z) + 2\pi(z), \qquad (5)$$

satisfies the condition $\tau'(z) < 0$, has a negative derivative, related to the function $\phi(s)$ by

$$\pi(z) = \sigma(z)\frac{d}{dz}[\ln \phi(z)]. \qquad (6)$$

The parameter $\lambda$ is defined by

$$\lambda = \lambda_n = -n\tau'(z) - \left[\frac{n(n-1)}{2}\sigma''\right]; \quad (n = 0,1,2,\ldots). \qquad (7)$$

The energy eigenvalues can be calculated from equation (7). In order to calculate the energy eigenvalues, we need first, to determine $\lambda$ by using the first derivative of $\pi(z)$ and defining

$$\lambda = k + \pi'(z). \qquad (8)$$

By solving the resulting quadratic equation for $\pi(z)$, we obtain the following expression

$$\pi(z) = \left(\frac{\sigma'-\bar{\tau}}{2}\right) \pm \sqrt{\left(\frac{\sigma'-\bar{\tau}}{2}\right)^2 - \bar{\sigma} + k\sigma}. \qquad (9)$$

Here, $\pi(z)$ is a polynomial with the parameter z and the prime denote the first derivative of the functions $\sigma(z)$ and $\tau(z)$, respectively. The determination of k is the essential point in the calculation of $\pi(z)$. It can be obtained by setting the discriminant of the square root to zero [37], therefore, a general quadratic expression for k can be obtained. On substitution of the values of $k, \pi'(z), \tau'(z)$ and $\sigma''$ into equations (7) and (8) and equating (7) and (8), one can evaluate the energy equation for any potential. The wave function $\phi(s)$ in equation (2) satisfies the condition

$$\frac{\phi'(z)}{\phi(z)} = \frac{\pi(z)}{\sigma(z)}, \qquad (10)$$

can be evaluated and using the Rodrigues' relation. The polynomial solutions $y_n(z)$ are given by

$$y_n(z) = \frac{C_n}{\rho(z)}\frac{d^n}{dz^n}[\sigma^n(z)\rho(z)], \qquad (11)$$

where $C_n$ is a normalization constant and the weight function $\rho(z)$ satisfies the following relation

$$\frac{d}{ds}[\sigma(z)\rho(z)] = \tau(z)\rho(z). \qquad (12)$$

A more generalized form of equation (2) for any potential may be presented as [39]:

$$\psi_{nl}''(z) + \left[\frac{\beta_1-\beta_2 z}{z(1-\beta_3 z)}\right]\psi_{nl}'(z) + \left[\frac{-\rho_2 z^2+\rho_1 z-\rho_0}{z^2(1-\beta_3)^2}\right]\psi_{nl}(z) = 0, \qquad (13)$$

which satisfies the wave functions of equation (3). By comparing equation (2) with equation (13), we have the following polynomials:

$$\bar{\tau}(z) = \beta_1 - \beta_2 z, \sigma(z) = z(1 - \beta_3 z), \bar{\sigma}(z) = -\rho_2 z^2 + \rho_1 z - \rho_0 . \tag{14}$$

Substituting equation (13) into equation (9), one obtains

$$\pi(z) = \beta_4 + \beta_5 z \pm \sqrt{(\beta_6 - k\beta_3)z^2 + (\beta_7 + k)z + \beta_8}, \tag{15}$$

with the following parametric constants:

$$\beta_4 = \frac{1}{2}(1 - \beta_1), \; \beta_5 = \frac{1}{2}(\beta_2 - 2\beta_3), \; \beta_6 = \beta_5^2 + \rho_2, \; \beta_7 = 2\beta_4\beta_5 - \rho_1, \; \beta_8 = \beta_4^2 + \rho_0. \tag{16}$$

According to Nikiforov-Uvarov method, the discriminant of equation (15) must be set to zero so that the expression for $k$ can be quadratically obtained as

$$k_{\pm} = -(\beta_7 + 2\beta_3\beta_8) \pm 2\sqrt{\beta_8\beta_9}, \; \beta_9 = \beta_3(\beta_7 + \beta_3\beta_8) + \beta_6 \tag{17}$$

Since the negative value of $k$ (that is $k_-$) gives the bound state solution, we consider

$$k_- = -(\beta_7 + 2\beta_3\beta_8) - 2\sqrt{\beta_8\beta_9}, \tag{18}$$

Inserting equation (18) into (15), we have

$$\pi(z) = \beta_4 + \beta_5 z - \left[\left(\sqrt{\beta_9} + \beta_3\sqrt{\beta_8}\right)z - \sqrt{\beta_8}\right], \tag{19}$$

having its first derivative as

$$\pi'(z) = \beta_5 - \left(\sqrt{\beta_9} + \beta_3\sqrt{\beta_8}\right) \tag{20}$$

Putting equations (14) and (15) into equation (5), one obtains

$$\tau(z) = \beta_1 + 2\beta_4 + (2\beta_5 - \beta_2)z - 2\left[\left(\sqrt{\beta_9} + \beta_3\sqrt{\beta_8}\right)z - \sqrt{\beta_8}\right], \tag{21}$$

and its first derivative becomes

$$\tau'(z) = -(\beta_2 - 2\beta_5) - 2\left[\left(\sqrt{\beta_9} + \beta_3\sqrt{\beta_8}\right)\right] . \tag{22}$$

By applying equation (16) in equation (22), we have

$$\tau'(z) = -2\beta_3 - 2\left[\left(\sqrt{\beta_9} + \beta_3\sqrt{\beta_8}\right)\right] < 0. \tag{23}$$

By applying equations (18) and (20) in equation (8), we have

$$\lambda = -(\beta_7 + 2\beta_3\beta_8) - 2\sqrt{\beta_8\beta_9} + \beta_5 - \left(\sqrt{\beta_9} + \beta_3\sqrt{\beta_8}\right). \tag{24}$$

With equations (14) and (22), the parameter $\lambda_n$ in equation (7) becomes

$$\lambda_n = \beta_2 n - 2n\beta_5 + 2n(\sqrt{\beta_9} + \beta_3\sqrt{\beta_8}) + n(n-1)\beta_3 ; \quad (n = 0, 1, 2, \ldots). \tag{25}$$

Equating equations (24) and (25), one obtains the bound state energy equation for any potential[37], [38], [39], [40] as:

$$\beta_2 n - (2n+1)\beta_5 + (2n+1)(\sqrt{\beta_9} + \beta_3\sqrt{\beta_8})$$
$$+n(n-1)\beta_3 + \beta_7 + 2\beta_3\beta_8 + 2\sqrt{\beta_8\beta_9} = 0, \tag{26}$$

Using equations (10), (11) and (12), the wave function parameters be evaluated as

$$\rho(z) = z^{\beta_{10}}(1-\beta_3 z)^{\beta_{11}}, \varphi(z) = z^{\beta_{12}}(1-\beta_3 z)^{\beta_{13}}, \beta_{12} > 0, \beta_{13} > 0,$$
$$y_{nl}(z) = P_{nl}^{(\beta_{10},\beta_{11})}(1 - 2\beta_3 z), \beta_{10} > -1, \beta_{11} > -1, \tag{27}$$

in such a way that the associated wave function in equation (3) becomes

$$\psi_{nl}(z) = N_{nl} z^{\beta_{12}}(1-\beta_3 z)^{\beta_{13}} P_{nl}^{(\beta_{10},\beta_{11})}(1 - 2\beta_3 z), \tag{28}$$

where $P_{nl}^{(\mu,\nu)}(x)$, $\mu > -1$, $\nu > -1$, $x\epsilon[-1, 1]$ are Jacobi polynomials with the following parametric constants;

$$\beta_{10} = \beta_1 + 2\beta_4 + 2\sqrt{\beta_8} - 1 > -1, \beta_{11} = \beta_2 - 2\beta_5 + 2(\sqrt{\beta_9} + \beta_3\sqrt{\beta_8}) > -1, \beta_3 \neq 0,$$

$$\beta_{12} = \beta_4 + \sqrt{\beta_8} > 0, \beta_{13} = \beta_5 - (\sqrt{\beta_9} - \beta_3\sqrt{\beta_8}), \beta_3 \neq 0, \tag{29}$$

By considering a special case where $\beta_3 = 0$, then the associated wave function reduces to the form [39], [40]:

$$\lim_{\beta_3 \to 0} P_{nl}^{(\beta_{10},\beta_{11})}(1 - 2\beta_3 z) = L_{nl}^{\beta_{10}}(\beta_{11}z), \lim_{\beta_3 \to 0}(1-\beta_3 z)^{\beta_{13}} = e^{\beta_{13}z},$$

$$\psi_{nl}(z) = N_{nl} z^{\beta_{12}} e^{\beta_{13}z} L_{nl}^{\beta_{10}}(\beta_{11}z), \tag{30}$$

where $L_{nl}^{\beta_{10}}(z)$ is well known as Laguerre polynomials.

## 3.0. Nonrelativistic eigensolutions of isotropic oscillator plus inverse quadratic potential in an external magnetic field

For a charged particle moving in a uniform magnetic field, the Hamiltonian of the system may be defined [16 and the references therein] as:

$$H = \frac{1}{2\mu}\left(\mathbf{p} + \frac{e}{c}\mathbf{A}\right)^2 + V(r), \tag{31}$$

where $m$, is the mass of the charged particle, $e$ is the electronic charge, $\mathbf{p}$ is the momentum of charged particle, $\mathbf{A} = \frac{1}{2}\mathbf{B}\times\mathbf{r}$ is the vector potential in the symmetric gauge, $c$ is the velocity of light and $V(r)$ is the cylindrical potential representing the potential in equation (1). The Hamiltonian for this system can be evaluated, in the CGS system and in atomic units $\hbar = e = 1$, as:

$$H = \frac{1}{2\mu}\left(-i\nabla + \frac{1}{2}\mathbf{B}\times\mathbf{r}\right)^2 + V(r), \tag{32}$$

and the Schrodinger equation yields:

$$H\varphi = \frac{1}{2\mu}\left(-i\nabla + \frac{1}{2}\mathbf{B}\times\mathbf{r}\right)^2\varphi + V(r)\varphi = i\partial_t\varphi = E\varphi, \tag{33}$$

Since this problem involves two-dimensions, therefore, it is sufficient enough to study in polar coordinates $(r,\phi)$ within the plane and to employ the following ansatz for the eigenfunction:

$$\varphi(r,\phi) = \frac{e^{im\phi}}{\sqrt{2\pi}}\frac{R(r)}{\sqrt{r}}, \quad m = 0, \pm 1, \pm 2 \ldots \tag{34}$$

Consequently, the radial wavefunction $R(r)$ must satisfy the following radial Schrödinger equation [16], [17]:

$$\frac{d^2R(r)}{dr^2} + 2[E - V_{eff}(r)]R(r) = 0, \tag{35}$$

with the effective potential $V_{eff}(r)$ defined as

$$V_{eff}(r) = m\omega_L + \frac{1}{2}\omega_L^2 r^2 + \frac{m^2 - \frac{1}{4}}{2r^2} + V(r), \tag{36}$$

where $\omega_L = B/2c$, $m$ and $E$ symbolizes the Larmor frequency, the eigenvalue of angular momentum and the energy spectra of the vibrating molecules respectively. By using the $V(r)$ as the isotropic oscillator plus inverse quadratic potential (IOPIQP), the effective potential influenced by an external magnetic field becomes

$$V_{eff}(r) = m\omega_L + \frac{\frac{m^2-\frac{1}{4}}{2}+g}{r^2} + \frac{1}{2}[\omega_L^2 + \mu\omega^2]r^2, \tag{37}$$

where $\mu$ and $\omega$ represent the mass and angular frequency of the vibrating molecules bounded by the IOPIQP, and the molecular constants for the selected diatomic molecules in this study are displayed in Table 2. Inserting equation (37) into equation (35) and apply a variable $z = r^2$, equation (35) can be transformation as

$$R''(z) + \left(\frac{1}{2z}\right)R'(z) + \left(\frac{-\rho_1 z^2 + \rho_2 z - \rho_3}{z^2}\right)R(z) = 0. \tag{38}$$

Comparing equation (38) with equation (13), we obtain the following analytical expressions:

$$\beta_1 = \frac{1}{2}, \beta_2 = \beta_3 = 0 \quad \rho_1 = \frac{\omega_L^2 + \mu\omega^2}{4}, \quad \rho_2 = \frac{E - m\omega_L}{2}, \quad \rho_3 = \frac{m^2 + 2g - \frac{1}{4}}{4}. \tag{39}$$

Using equations (16) and (29), other values of parametric constants $\beta_i (i = 4, 5, 6, \ldots)$ and their analytical values require for the derivation of energy eigenvalues and eigenfunctions are obtained and displayed in Table 1 as:

**Table 1**: The values of parametric constants require for the derivation of energy eigenvalues and eigenfunctions

| Parametric constants | Analytical values |
|---|---|
| $\beta_4$ | ¼ |
| $\beta_5$ | 0 |
| $\beta_6$ | $\frac{\omega_L^2 + \mu\omega^2}{4}$ |
| $\beta_7$ | $\frac{m\omega_L - E}{2}$ |
| $\beta_8$ | $\frac{m^2 + 2g}{4}$ |
| $\beta_9$ | $\frac{\omega_L^2 + \mu\omega^2}{4}$ |
| $\beta_{10}$ | $\sqrt{m^2 + 2g} > -1$ |
| $\beta_{11}$ | $\sqrt{\omega_L^2 + \mu\omega^2}$ |
| $\beta_{12}$ | $\frac{1 + 2\sqrt{m^2 + 2g}}{4} > 0$ |
| $\beta_{13}$ | $-\sqrt{\frac{\omega_L^2 + \mu\omega^2}{4}}$ |

Using the analytical values in Table 1 for the parametric constants $\beta_i (i = 1, 2, 3, \ldots)$ in equations (26) and (30), the energy eigenvalues and the normalized radial eigenfunctions for the IOPIQP in the presence of external magnetic field are obtain respectively as

$$E = m\omega_L + \sqrt{\omega_L^2 + \mu\omega^2}\left(2n + 1 + \sqrt{m^2 + 2g}\right), \tag{40}$$

$$R(r) = \left[\frac{2n!\gamma^{2\delta+2}}{(n+2\delta+1)!}\right]^{1/2} r^{2\delta+3/2} e^{-\frac{1}{2}\gamma r^2} L_n^{2\delta+1}(\gamma r^2), \tag{41}$$

where $\delta = -\frac{1}{2} + \sqrt{\frac{m^2}{4} + \frac{g}{2}}$, $\gamma = \sqrt{\omega_L^2 + \mu\omega^2}$ and $L_n^{2\delta+1}(\gamma r^2)$ is the associated Laguerre polynomial.

### 3.1. Expectation values ($r^{-2}$, $p^2$, T, and V) of isotropic oscillator plus inverse quadratic potential in an external magnetic field

The Hellmann-Feynman Theorem (HFT) is one of the useful techniques for obtaining expectation values of some quantum mechanical observables for any arbitrary values of quantum numbers [41], [42]. Suppose the Hamiltonian $H(\alpha)$ for a particular quantum mechanical system depends parameter $\alpha$ such that $E_{nl}(\alpha)$ and $\psi_{nm}(\alpha)$ are the eigenvalues and the eigenfunctions respectively. Therefore, the Hellmann-Feynman Theorem (HFT) states that:

$$\frac{\partial E_{nm}(\alpha)}{\partial \alpha} = \langle \psi_{nm}(\alpha) \left| \frac{\partial H(\alpha)}{\partial \alpha} \right| \psi_{nm}(\alpha) \rangle, \tag{42}$$

provided that the normalized eigenfunctions $\psi_{nm}(\alpha)$ are continuous, differentiable with respect to parameter $\alpha$. The effective Hamiltonian of isotropic oscillator plus inverse quadratic potential in an external magnetic field is given as

$$H = -\frac{1}{2}\frac{d^2}{dr^2} + m\omega_L + \frac{\frac{m^2 - \frac{1}{4}}{2} + g}{r^2} + \frac{1}{2}[\omega_L^2 + \mu\omega^2]r^2. \tag{43}$$

To find the expectation value of $r^2$, we let $\alpha = \omega$ such that equation (42) becomes

$$\frac{\partial E_{nm}(\omega)}{\partial \omega} = \langle \psi_n(\omega) \left| \frac{\partial H(\omega)}{\partial \omega} \right| \psi_n(\omega) \rangle. \tag{44}$$

Taking the first derivative of the effective Hamiltonian $H(\omega)$ in equation (43) with respect to vibrational frequency $\omega$, one obtains

$$\frac{\partial H(\omega)}{\partial \omega} = \mu\omega\langle r^2 \rangle. \tag{45}$$

The first derivative of the energy eigenvalues $E(\omega)$ in equation (40) with respect to vibrational frequency $\omega$, we have

$$\frac{\partial E_n m(\omega)}{\partial \omega} = \frac{\mu\omega\left(2n+1+\sqrt{m^2+2g}\right)}{\sqrt{\omega_L^2+\mu\omega^2}}. \tag{46}$$

With equations (45) and (46) in equation (44), it is easy for one to evaluate $\langle r^2 \rangle$ as:

$$\langle r^2 \rangle = \frac{(2n+1+\sqrt{m^2+2g})}{\sqrt{\omega_L^2+\mu\omega^2}}. \tag{47}$$

In principle and with $\alpha = \mu$ in equation (44), the expectation values of $p^2$, T, and V, can be obtained respectively, as

$$\langle p^2 \rangle = -\frac{\mu^2\omega^2(2n+1+\sqrt{m^2+2g})}{\sqrt{\omega_L^2+\mu\omega^2}}, \tag{48}$$

$$\langle T \rangle = -\frac{\mu\omega^2(2n+1+\sqrt{m^2+2g})}{2\sqrt{\omega_L^2+\mu\omega^2}}, \tag{49}$$

$$\langle V \rangle = \left[ m\omega_L + \sqrt{\omega_L^2 + \mu\omega^2}\left(2n + 1 + \sqrt{m^2 + 2g}\right) + \frac{\frac{\mu\omega^2}{2}(2n+1+\sqrt{m^2+2g})}{\sqrt{\omega_L^2+\mu\omega^2}} \right]. \tag{50}$$

**3.2. Diamagnetic susceptibility of isotropic oscillator plus inverse quadratic potential in an external magnetic field**

The diamagnetic susceptibility is given [3], [21] as

$$\chi = -\frac{Nze^2}{6\mu c^2}\langle r^2 \rangle, \tag{51}$$

where N is the Avogadro number, z is the atomic number, e is the electronic charge, c is the speed of light, $\mu$ is the effective mass of the vibrating molecules in this study. It has been found that diamagnetism is a fundamental magnetic phenomenon that explains the tendency of electric charges to partially shield the interior of a body from external magnetic field and that diamagnetic materials possess magnetic effects due to an external field which alters electron motion within the atoms [21]. Using equation (47) in (51) defines the diamagnetic susceptibility for the Isotropic oscillator plus Inverse quadratic potential in an external magnetic field as

$$\chi = -\frac{Nze^2}{6\mu c^2}\left[\frac{(2n+1+\sqrt{m^2+2g})}{\sqrt{\omega_L^2+\mu\omega^2}}\right], \tag{52}$$

where all the symbols have been explained accordingly. The corresponding magnetic moment $\mu_B$ can be expressed as

$$\mu_B = -\frac{e^2}{6\mu c^2}\langle r^2\rangle B = -\frac{2e^2\omega_L}{6\mu c}\left[\frac{(2n+1+\sqrt{m^2+2g})}{\sqrt{\omega_L^2+\mu\omega^2}}\right]. \tag{53}$$

## 4. Results and Discussions

In order to verify the reliability, validity and consistency of our results, using the molecular constants in Table 2 [21], we present the computed results for the energy spectra, expectation values $\langle r^2\rangle, \langle p^2\rangle, \langle T\rangle, \langle V\rangle$ and diamagnetic susceptibility $\chi$ for the selected molecules bounded by Isotropic oscillator plus Inverse quadratic potential with varying Larmor frequencies $\omega_L$ for the case of low vibrational energy level ($n = 0, 1, 2, 3$), low rotational energy level ($m = 0, +1$) and inverse quadratic potential strength ($g = 0, 1$) in Tables 3-11. Figure 1-4 shows the variation of expectation values for some quantum mechanical observables as function of Larmor frequencies $\omega_L$ for various rotational energy level ($n = 0, 1, 2, 3$) with $g = m = 1$ for the selected molecules in Table 2. Figure 5 shows the variation of diamagnetic susceptibility as function of Larmor frequencies $\omega_L$ for various rotational energy level $n$ with $g = m = z = e = 1$ for the molecules in Table 2. In all the calculations, we have also employed the following recently used conversions $1\ a.m.u = 1.66 \times 10^{-27} kg$, $c = 3.00 \times 10^8 m/s$ and $N = 6.02 \times 10^{23}$ moles. All our results are in their standard units.

**Table 2**: Molecular Parameters for this Study [21]

| Molecules | Vibrational frequencies $\omega \times 10^{13} s^{-1}$ | Mass $\mu$ in a.m.u |
|---|---|---|
| CO | 6.471 | 6.8606719 |
| HCl | 8.814 | 0.9801045 |
| $I_2$ | 0.642 | 63.45223502 |
| $H_2$ | 12.960 | 0.50391 |

It was observed that energy spectra increases with increase in magnetic field strength (Larmor frequency), vibrational level, and inverse quadratic potential strength $g$ for all the selected molecules. See Tables 3-6. This observation suggest that the energy spectra of the selected molecules would be affected significantly by the external magnetic field. The expectation value $\langle r^2\rangle$ which are all positive increases with increasing rotational energy level $n$ but decreases monotonically towards zero with increase in magnetic field strengths for all the selected molecules for $g = m = 1$. See Table 7 and Figure 1. The expectation value $\langle p^2\rangle$ which are all negative decreases with increasing rotational energy level $n$ but increases with increase in magnetic field strengths and tend to converge at a very high magnetic field strength (Larmor frequencies $\omega_L > 10$) for all the selected molecules for $g = m = 1$. See Table 8 and Figure 2.

The expectation value $\langle T \rangle$ which are all negatives decreases with increasing rotational energy level $n$ but increases monotonically with increase in magnetic field strengths and tend to converge at a very high magnetic field strength (Larmor frequencies $\omega_L > 10$) for all the selected molecules for $g = m = 1$. See Table 9 and Figure 3. The expectation value $\langle V \rangle$ which are all positive increases with increasing rotational energy level $n$ as well as the magnetic field strength. A clear divergence is noticeable at all values of magnetic field strength for all the selected molecules for $g = m = 1$. See Table 10 and Figure 4.

The diamagnetic susceptibility $X$ which are all negatives increases monotonically with increasing rotational energy level $n$ as well as the magnetic field strength and tend to converge at any $\omega_L > 10$ for all the selected molecules for $g = m = z = e = 1$. See Table 11 and Figure 5.

**Table 3**: Energy spectra for CO molecules bounded by Isotropic oscillator plus Inverse quadratic potential for arbitrary Larmor frequencies $\omega_L$

| $n$ | $m = 0$, $\omega_L = 0$ | $m = 1$, $\omega_L = 0$ | $m = 0$, $\omega_L = 5$ | $m = 1$, $\omega_L = 5$ | $m = 0$, $\omega_L = 10$ | $m = 1$, $\omega_L = 10$ |
|---|---|---|---|---|---|---|
| | | | | $g = 0$ | | |
| 0 | 6.90572 | 13.8114 | 8.52578 | 22.0516 | 12.1527 | 34.3055 |
| 1 | 20.7172 | 27.6229 | 25.5773 | 39.1031 | 36.4582 | 58.6109 |
| 2 | 34.5286 | 41.4343 | 42.6289 | 56.1547 | 60.7637 | 82.9164 |
| 3 | 48.3488 | 55.2457 | 59.6805 | 73.2062 | 85.0691 | 107.222 |
| | | | | $g = 1$ | | |
| 0 | 16.6719 | 18.8668 | 20.5831 | 28.2929 | 29.3393 | 43.2019 |
| 1 | 30.4833 | 32.6782 | 37.6346 | 45.3444 | 53.6448 | 67.5074 |
| 2 | 44.2947 | 46.4896 | 54.6862 | 62.3960 | 77.9502 | 91.8128 |
| 3 | 58.1062 | 60.3011 | 71.7377 | 79.4475 | 102.256 | 116.118 |

**Table 4**: Energy spectra for HCl molecules bounded by Isotropic oscillator plus Inverse quadratic potential for arbitrary Larmor frequencies $\omega_L$

| $n$ | $m = 0$, $\omega_L = 0$ | $m = 1$, $\omega_L = 0$ | $m = 0$, $\omega_L = 5$ | $m = 1$, $\omega_L = 5$ | $m = 0$, $\omega_L = 10$ | $m = 1$, $\omega_L = 10$ |
|---|---|---|---|---|---|---|
| | | | $g = 0$ | | | |
| 0 | 3.55519 | 7.11039 | 6.1351 | 17.2702 | 10.6132 | 31.2263 |
| 1 | 10.6656 | 14.2208 | 18.4053 | 29.5404 | 31.8395 | 52.4527 |
| 2 | 17.7760 | 21.3312 | 30.6755 | 41.8106 | 53.0659 | 73.6790 |
| 3 | 24.8864 | 28.4416 | 42.9457 | 54.0808 | 74.2922 | 94.9054 |
| | | | $g = 1$ | | | |
| 0 | 8.5830 | 9.71297 | 14.8114 | 21.7614 | 25.6225 | 38.9957 |
| 1 | 15.6934 | 16.8234 | 27.0816 | 34.0316 | 46.8488 | 60.2221 |
| 2 | 22.8038 | 23.9337 | 39.3518 | 46.3018 | 68.0751 | 81.4484 |
| 3 | 29.9142 | 31.0441 | 51.6220 | 58.5720 | 89.3015 | 102.675 |

**Table 5**: Energy spectra for $I_2$ molecules bounded by Isotropic oscillator plus Inverse quadratic potential for arbitrary Larmor frequencies $\omega_L$

| $n$ | $m = 0$, $\omega_L = 0$ | $m = 1$, $\omega_L = 0$ | $m = 0$, $\omega_L = 5$ | $m = 1$, $\omega_L = 5$ | $m = 0$, $\omega_L = 10$ | $m = 1$, $\omega_L = 10$ |
|---|---|---|---|---|---|---|
| | | | $g = 0$ | | | |
| 0 | 2.08359 | 4.16718 | 5.41677 | 15.8335 | 10.2148 | 30.4295 |
| 1 | 6.25077 | 8.33437 | 16.2503 | 26.6671 | 30.6443 | 50.859 |
| 2 | 10.4180 | 12.5015 | 27.0838 | 37.5006 | 51.0738 | 71.2886 |
| 3 | 14.5851 | 16.6687 | 37.9174 | 48.3341 | 71.5033 | 91.7181 |
| | | | $g = 1$ | | | |
| 0 | 5.03023 | 5.69248 | 13.0772 | 19.7989 | 24.6606 | 37.9072 |
| 1 | 9.19742 | 9.85966 | 23.9108 | 30.6324 | 45.0901 | 58.3368 |
| 2 | 13.3646 | 14.0268 | 34.7443 | 41.4659 | 65.5197 | 78.7663 |
| 3 | 17.5318 | 18.1940 | 45.5778 | 52.2995 | 85.9492 | 99.1958 |

**Table 6**: Energy spectra for $H_2$ molecules bounded by Isotropic oscillator plus Inverse quadratic potential for arbitrary Larmor frequencies $\omega_L$

| $n$ | $m = 0$, $\omega_L = 0$ | $m = 1$, $\omega_L = 0$ | $m = 0$, $\omega_L = 5$ | $m = 1$, $\omega_L = 5$ | $m = 0$, $\omega_L = 10$ | $m = 1$, $\omega_L = 10$ |
|---|---|---|---|---|---|---|
| | | | $g = 0$ | | | |
| 0 | 3.74831 | 7.49662 | 6.24899 | 17.4980 | 10.6794 | 31.3588 |
| 1 | 11.2449 | 14.9932 | 18.7470 | 29.9959 | 32.0382 | 52.7176 |
| 2 | 18.7416 | 22.4899 | 31.2449 | 42.4939 | 53.3971 | 74.0765 |
| 3 | 26.2382 | 29.9865 | 43.7429 | 54.9919 | 74.7559 | 95.4353 |
| | | | $g = 1$ | | | |
| 0 | 9.04922 | 10.2406 | 15.0864 | 22.0725 | 25.7824 | 39.1767 |
| 1 | 16.5458 | 17.7372 | 27.5844 | 34.5705 | 47.1412 | 60.5355 |
| 2 | 24.0425 | 25.2338 | 40.0823 | 47.0685 | 68.5000 | 81.8943 |
| 3 | 31.5391 | 32.7304 | 52.5803 | 59.5665 | 89.8588 | 103.253 |

**Table 7**: Expectation values $\langle r^2 \rangle$ for CO, HCl, $I_2$ and $H_2$ molecules bounded by Isotropic oscillator plus Inverse quadratic potential for arbitrary Larmor frequencies $\omega_L$ with $g = m = 1$.

| | CO | | | HCl | | |
|---|---|---|---|---|---|---|
| $n$ | $\omega_L = 0$ | $\omega_L = 5$ | $\omega_L = 10$ | $\omega_L = 0$ | $\omega_L = 5$ | $\omega_L = 10$ |
| 0 | 0.395622 | 0.320446 | 0.224810 | 0.768467 | 0.445315 | 0.257421 |
| 1 | 0.685237 | 0.555028 | 0.389382 | 1.331020 | 0.771308 | 0.445866 |
| 2 | 0.974852 | 0.789611 | 0.553954 | 1.893580 | 1.097300 | 0.634311 |
| 3 | 1.264470 | 1.024190 | 0.718526 | 2.456140 | 1.423290 | 0.822756 |
| | $I_2$ | | | $H_2$ | | |
| 0 | 1.31122 | 0.504369 | 0.267461 | 0.728875 | 0.437199 | 0.255824 |
| 1 | 2.27110 | 0.873593 | 0.463256 | 1.262450 | 0.757251 | 0.443100 |
| 2 | 3.23098 | 1.242820 | 0.659051 | 1.796020 | 1.077300 | 0.630377 |
| 3 | 4.19087 | 1.612040 | 0.854846 | 2.329600 | 1.397350 | 0.817653 |

**Table 8**: Expectation values $\langle \bar{p}^2 \rangle$ for CO, HCl, $I_2$ and $H_2$ molecules bounded by Isotropic oscillator plus Inverse quadratic potential for arbitrary Larmor frequencies $\omega_L$ with $g = m = 1$.

| | CO | | | HCl | | |
|---|---|---|---|---|---|---|
| $n$ | $\omega_L = 0$ | $\omega_L = 5$ | $\omega_L = 10$ | $\omega_L = 0$ | $\omega_L = 5$ | $\omega_L = 10$ |
| 0 | -2.14868e-25 | -1.74039e-25 | -1.22098e-25 | -1.58027e-26 | -9.15745e-27 | -5.29359e-27 |
| 1 | -3.72163e-25 | -3.01445e-25 | -2.11479e-25 | -2.73712e-26 | -1.58612e-26 | -9.16877e-27 |
| 2 | -5.29457e-25 | -4.28850e-25 | -3.00861e-25 | -3.89396e-26 | -2.25649e-26 | -1.30440e-26 |
| 3 | -6.86752e-25 | -5.56256e-25 | -3.90242e-25 | -5.0508e-26 | -2.92686e-26 | -1.69191e-26 |
| | $I_2$ | | | $H_2$ | | |
| 0 | -5.99593e-25 | -2.30637e-25 | -1.22304e-25 | -8.56614e-27 | -5.13820e-27 | -3.00659e-27 |
| 1 | -1.03852e-24 | -3.99475e-25 | -2.11837e-25 | -1.48370e-26 | -8.89963e-27 | -5.20756e-27 |
| 2 | -1.47746e-24 | -5.68313e-25 | -3.01369e-25 | -2.11078e-26 | -1.26611e-26 | -7.40853e-27 |
| 3 | -1.91639e-24 | -7.37151e-25 | -3.90902e-25 | -2.73787e-26 | -1.64225e-26 | -9.60951e-27 |

**Table 9**: Expectation values $\langle \bar{T} \rangle$ for CO, HCl, $I_2$ and $H_2$ molecules bounded by Isotropic oscillator plus Inverse quadratic potential for arbitrary Larmor frequencies $\omega_L$ with $g = m = 1$

| | CO | | | HCl | | |
|---|---|---|---|---|---|---|
| $n$ | $\omega_L = 0$ | $\omega_L = 5$ | $\omega_L = 10$ | $\omega_L = 0$ | $\omega_L = 5$ | $\omega_L = 10$ |
| 0 | -9.43338 | -7.64086 | -5.36046 | -4.85649 | -2.81426 | -1.62682 |
| 1 | -16.3391 | -13.2344 | -9.28460 | -8.41168 | -4.87444 | -2.81774 |
| 2 | -23.2448 | -18.8279 | -13.2087 | -11.9669 | -6.93462 | -4.00866 |
| 3 | -30.1505 | -24.4214 | -17.1329 | -15.5221 | -8.99480 | -5.19957 |
| | $I_2$ | | | $H_2$ | | |
| 0 | -2.84624 | -1.09482 | -0.580571 | -5.12029 | -3.07129 | -1.79714 |
| 1 | -4.92983 | -1.89629 | -1.00558 | -8.86860 | -5.31962 | -3.11274 |
| 2 | -7.01342 | -2.69775 | -1.43059 | -12.6169 | -7.56796 | -4.42834 |
| 3 | -9.09701 | -3.49922 | -1.85559 | -16.3652 | -9.81600 | -5.74390 |

**Table 10**: Expectation values $\langle \bar{V} \rangle$ for CO, HCl, $I_2$ and $H_2$ molecules bounded by Isotropic oscillator plus Inverse quadratic potential for arbitrary Larmor frequencies $\omega_L$ with $g = m = 1$.

| | CO | | | HCl | | |
|---|---|---|---|---|---|---|
| $n$ | $\omega_L = 0$ | $\omega_L = 5$ | $\omega_L = 10$ | $\omega_L = 0$ | $\omega_L = 5$ | $\omega_L = 10$ |
| 0 | 28.3002 | 35.9337 | 48.5623 | 14.5695 | 24.5757 | 40.6225 |
| 1 | 49.0173 | 58.5788 | 76.7919 | 25.2350 | 38.9060 | 63.0398 |
| 2 | 69.7345 | 81.2238 | 105.022 | 35.9006 | 53.2364 | 85.4571 |
| 3 | 90.4516 | 103.869 | 133.251 | 46.5662 | 67.5668 | 107.874 |
| | $I_2$ | | | $H_2$ | | |
| 0 | 8.53872 | 20.8937 | 38.4878 | 15.3609 | 25.1438 | 40.9738 |
| 1 | 14.7895 | 32.5287 | 59.3423 | 26.6058 | 39.8901 | 63.6483 |
| 2 | 21.0403 | 44.1637 | 80.1969 | 37.8507 | 54.6365 | 86.3227 |
| 3 | 27.2910 | 55.7987 | 101.051 | 49.0957 | 69.3828 | 108.997 |

**Table 11**: Diamagnetic Susceptibilities $X$ for CO, HCl, $I_2$ and $H_2$ molecules bounded by Isotropic oscillator plus Inverse quadratic potential for arbitrary Larmor frequencies $\omega_L$ with $g = m = z = e = 1$.

| | CO | | | HCl | | |
|---|---|---|---|---|---|---|
| $n$ | $\omega_L = 0$ | $\omega_L = 5$ | $\omega_L = 10$ | $\omega_L = 0$ | $\omega_L = 5$ | $\omega_L = 10$ |
| 0 | -3.87265e+31 | -3.13677e+31 | -2.20061e+31 | -5.26560e+32 | -3.05133e+32 | -1.76387e+32 |
| 1 | -6.70762e+31 | -5.43304e+31 | -3.81157e+31 | -9.12028e+32 | -5.28506e+32 | -3.05511e+32 |
| 2 | -9.54260e+31 | -7.72932e+31 | -5.42252e+31 | -1.29750e+33 | -7.51880e+32 | -4.34635e+32 |
| 3 | -1.23776e+32 | -1.00256e+32 | -7.03348e+31 | -1.68297e+33 | -9.75253e+32 | -5.63759e+32 |
| | $I_2$ | | | $H_2$ | | |
| 0 | -1.38779e+31 | -5.33822e+30 | -2.83079e+30 | -9.71393e+32 | -5.82668e+32 | -3.40944e+32 |
| 1 | -2.40372e+31 | -9.24607e+30 | -4.90308e+30 | -1.68250e+33 | -1.00921e+33 | -5.90532e+32 |
| 2 | -3.41966e+31 | -1.31539e+31 | -6.97536e+30 | -2.39361e+33 | -1.43575e+33 | -8.40121e+32 |
| 3 | -4.43559e+31 | -1.70618e+31 | -9.04765e+30 | -3.10472e+33 | -1.86229e+33 | -1.08971e+33 |

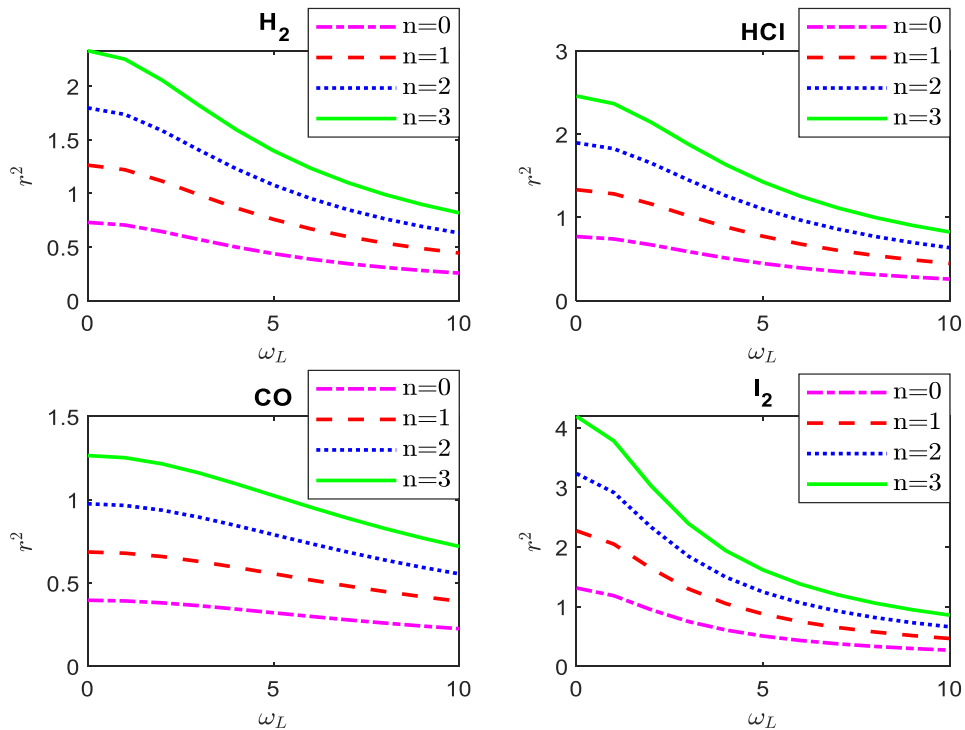

**Figure 1**: The expectation value of the square of position $r^2$ as a function of Larmor frequencies $\omega_L$ for various number of states $n$ with $g = m = 1$.

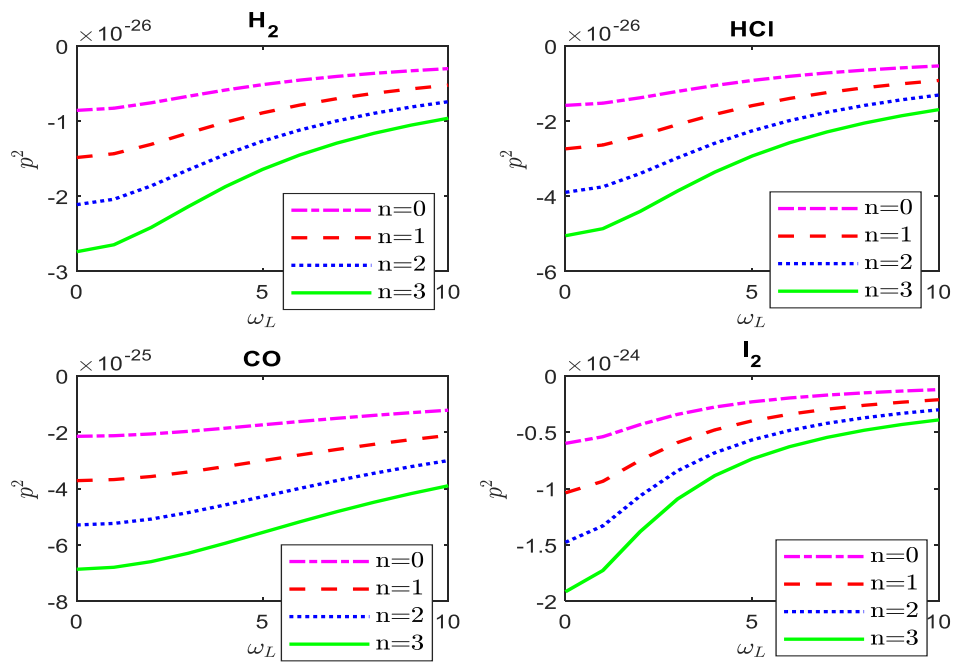

**Figure 2**: The expectation value of the square of momentum $p^2$ as a function of Larmor frequencies $\omega_L$ for the molecules ($H_2, HCl, CO, I_2$) and for various number of states $n$ with $g = m = 1$.

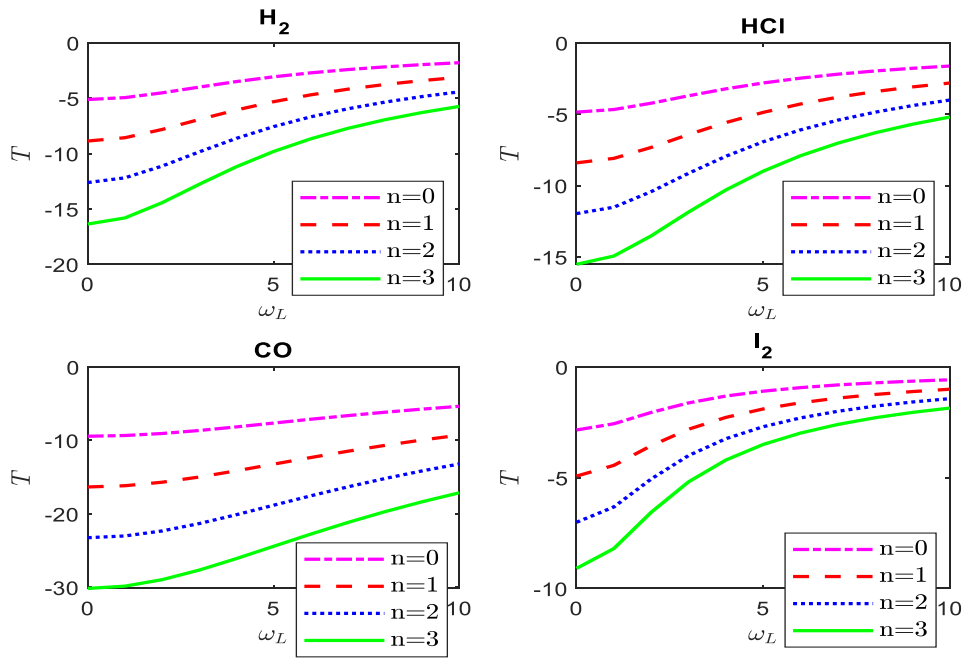

**Figure 3**: The expectation value of kinetic energy $T$ as a function of Larmor frequencies $\omega_L$ for various number of states $n$ with $g = m = 1$.

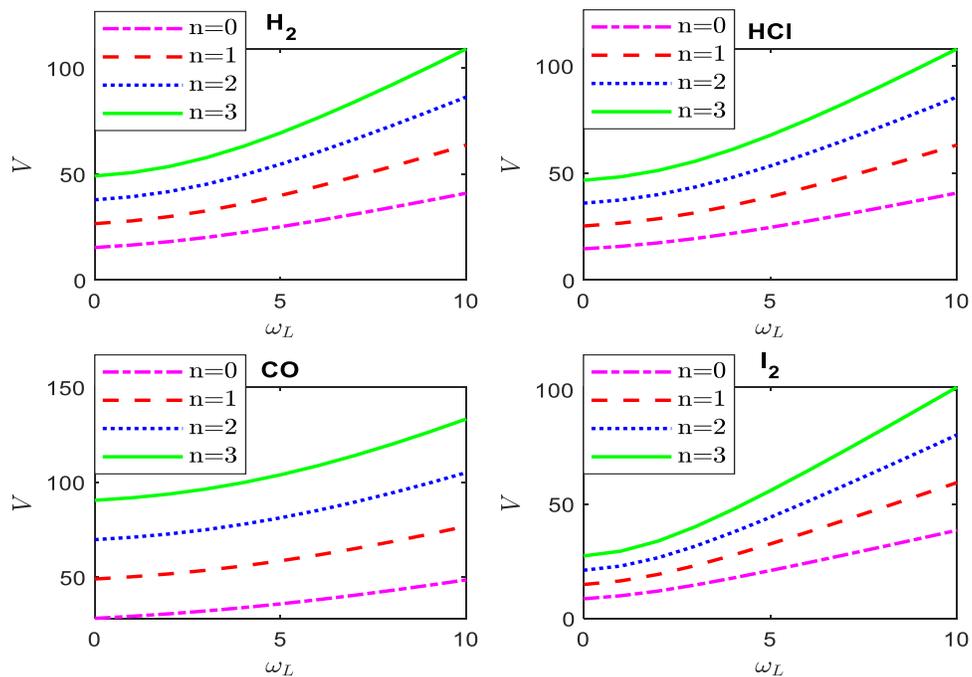

**Figure 4**: The expectation value of potential energy $V$ as a function of Larmor frequencies $\omega_L$ for the molecules ($H_2, HCl, CO, I_2$) and for various number of states $n$ with $g = m = 1$.

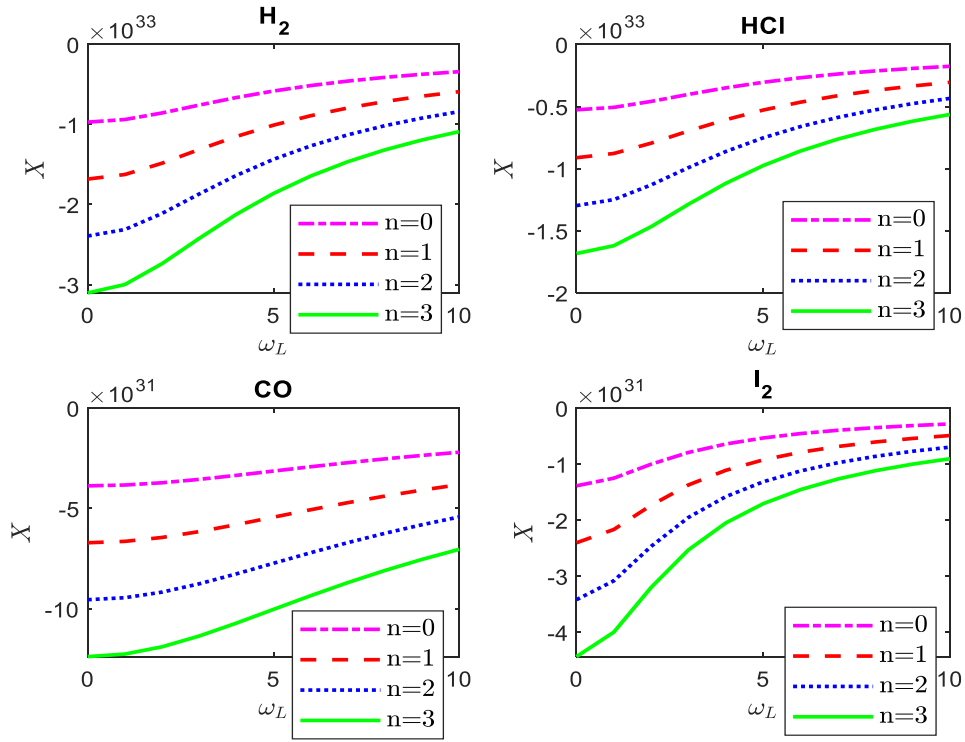

**Figure 5**: The Diamagnetic Susceptibilities $X$ as a function of Larmor frequencies $\omega_L$ for the molecules ($H_2, HCl, CO, I_2$) and various number of states $n$ with $g = m = z = e = 1$.

## 5. Concluding Remarks

We have studied the two dimensional radial Schrödinger equation with isotropic harmonic oscillator plus inverse quadratic potential in an external magnetic field via parametric Nikiforov-Uvarov method. Energy eigenvalues equation, normalized wavefunction, expressions for expectation values square of position $\langle r^2 \rangle$, square of momentum $\langle p^2 \rangle$, kinetic energy $\langle T \rangle$, potential energy $\langle V \rangle$ and diamagnetic susceptibility $X$ for the interaction potential model have been obtained. The computed results for the energy spectra, expectation values $\langle r^2 \rangle, \langle p^2 \rangle, \langle T \rangle, \langle V \rangle$ and diamagnetic susceptibility $X$ for some diatomic molecules bounded by the isotropic harmonic oscillator plus inverse quadratic potential for low vibrational and rotational levels are found to have strong dependence on the magnetic field strengths as well as the inverse quadratic potential strength $g$.

The expectation values $\langle r^2 \rangle, \langle p^2 \rangle, \langle T \rangle$ and diamagnetic susceptibility $X$ as a function of Larmor frequencies $\omega_L$ for all the molecules tend to converge at a very high magnetic field strength ($\omega_L > 10$). This may signifies a case where low vibrational and rotational energy levels cease to have any significant effect on the expectation values of $\langle r^2 \rangle, \langle p^2 \rangle$ and $\langle T \rangle$ and diamagnetic susceptibility $X$ for all the molecules, despite the increasing magnetic field strength. A divergence is noticeable for the expectation value of potential energy $\langle V \rangle$ which signifies that the rotational and vibration energy levels will continue to have meaningful effect as long as the magnetic field strength increases. Also as the magnetic field strength increases, diamagnetic susceptibility $X$ increases, which may likely increase the tendency of magnetic field to shield or alter the vibrational motion of the molecules bounded by the potential model in this study.

Using the results obtained in this work one can study the thermodynamic properties of the system, Shannon entropy and Fisher information for the first and second excited states. And this would be our focus in the subsequent research work.

**Data Availability Statement**

All the data used are included in this paper

**Declarations of Interest Statement**

**Ethics approval and consent to participate**

Not applicable

**Consent for publication**

Not applicable

**Competing interests**

The authors declared that there is no competing interests regarding the paper.

**Funding**

No funding is received

**Acknowledgements**

The authors wish to thank some authors for making their papers available for development of this paper. The authors also wish to thank all the referees for their invaluable comments with regard to this paper.